\documentclass[aps,prl,twocolumn]{revtex4} 
\usepackage{amssymb}
\usepackage{graphicx} 
\usepackage{amsmath}
\usepackage{color}

\begin{document} 
\title{Large Surface Magnetization in Noncentrosymmetric Antiferromagnets} 

\author{Mike A. Lund$^1$, Karin Everschor-Sitte$^2$, and Kjetil M. D.\ Hals$^1$} 
\affiliation{$^1$ Department of Engineering Sciences, University of Agder, 4879 Grimstad, Norway\\
$^2$ Institute of Physics, Johannes Gutenberg Universit\"at, 55128 Mainz, Germany} 
%%%%%%%%%%%%%%%%%%%%%%%%%%%%%%%%%%%%%%%%%%
\newcommand{\Kjetil}[1]{\textcolor{red}{#1}} 
\newcommand{\Karin}[1]{\textcolor{green}{#1}}
%%%%%%%%%%%%%%%%%%%%%%%%%%%%%%%%%%%%%%%%%%%%%%%%%%%%%%%%%%%%%%%%%%%%%%%%%%%%%%% 
\begin{abstract}
Thin-film antiferromagnets (AFs) with Rashba spin-orbit coupling are theoretically investigated. We demonstrate that the relativistic Dzyaloshinskii-Moriya interaction (DMI) produces a large surface magnetization and a boundary-driven twist state in the antiferromagnetic N\' eel vector. We predict a magnetization on the order of $2.3\cdot 10^4$~A/m, which is comparable to the magnetization of ferromagnetic semiconductors. 
Importantly, the magnetization is characterized by ultra-fast terahertz dynamics and provides new approaches for efficiently probing and controlling the spin dynamics of AFs as well as detecting the antiferromagnetic DMI.
Notably, the magnetization does not lead to any stray magnetic fields except at corners where weak magnetic monopole fields appear.
\end{abstract}

\maketitle 

%%%%%%%%%%%%%%%%%%%%%%%%%%%%%%%%%%%%%%%%%%%%%%%%%%%%%%%%%%%%%%%%%%%%%%%%%%%%%%% 
% Introduction: 
%%%%%%%%%%%%%%%%%%%%%%%%%%%%%%%%%%%%%%%%%%%%%%%%%%%%%%%%%%%%%%%%%%%%%%%%%%%%%%% 
Spintronic devices have traditionally been based on ferromagnets~\cite{Chappert:nm2007,Ralph:nm2007,Brataas:nm2012}. However, over the last years a new and promising direction within the field of spin-based electronics has emerged, which exploits the unique physical properties of antiferromagnets (AFs) for developing new ultra-fast information technologies~\cite{Jungwirth:np2018,Duine:np2018,Gomonay:np2018,Zelezny:np2018,Nemec:np2018,Libor:np2018}. The AFs are ordered spin systems, in which the direction of the localized magnetic moments alternate between neighboring lattice sites in such a way that the net magnetization vanishes in equilibrium. Two of the main arguments for using AFs in spin electronics are their terahertz (THz) spin dynamics, which is a thousand times faster than ferromagnets, and the absence of stray magnetic fields that severely limit the density of ferromagnetic bits in magnetoresistive random-access memories. The combined effect of zero stray fields and ultra-fast spin dynamics implies that AFs could pave the way for information storage systems with considerably improved density, stability, and writing speed of the magnetic bits~\cite{Jungwirth:np2018}.   

However, the zero net magnetization of AFs also leads to disadvantages relative to ferromagnets. Firstly, it makes the staggered spin order nearly invisible on the outside. Consequently, it is extremely difficult to detect the spin dynamics of AFs and probe the switching of the antiferromagnetic order parameter. Secondly, the staggered spin order weakly couples to uniform spin densities and is thus hard to control and manipulate via spin-polarized currents.
While recent works have demonstrated that the antiferromagnetic spin order can be manipulated using electric currents~\cite{Wadley:science2016,Reichlova:prb2015,Nunez:prb2006,Duine:prb2007,Gomonay:jmj2008,Wang:prl2008,Haney:prl2008,Gomonay:prb2010,Hals:prl2011,Manchon:prb2014,Cheng:prb2014,Cheng:prl2014,Velkov:njp2016} as well as laser pump pulses~\cite{Duong:prl2004,Kimel:n2004,Manz:np2016}, these are applicable only in systems with engineered symmetry properties. This work demonstrates that these major challenges can be overcome generically in noncentrosymmetric thin-film AFs with strong spin-orbit coupling (SOC), because here a significant surface magnetization with THz dynamics emerges.

\begin{figure}[ht] 
\centering 
\includegraphics[scale=1.0]{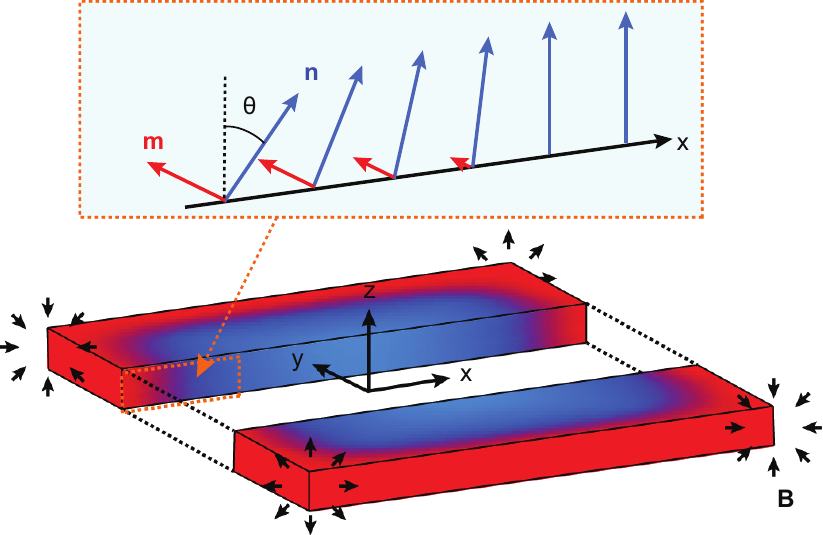}  
\caption{(color online). Sketch of an AF with DMI.
In the bulk, the antiferromagnetic order parameter $\boldsymbol{n}$ is aligned along the easy axis (i.e., the $z$-axis). At the boundaries, the DMI produces a twist state in $\boldsymbol{n}$, which is accompanied by a large surface magnetization $\boldsymbol{m}$. The magnetization is aligned along the surface (along $y$ in the inset). Therefore, it does not lead to any surface poles and stray fields, except at the corners of the sample where monopole fields are generated. Importantly, the strength of the stray field $\boldsymbol{B}$ can be tuned via the sample geometry and vanishes for quadratic thin-films.}
\label{Fig1} 
\end{figure} 

In noncentrosymmetric magnets, the SOC produces a spatially asymmetric exchange interaction, which is commonly referred to as the Dzyaloshinskii-Moriya interaction (DMI)~\cite{Dzyaloshinsky:jpcs1958,Moriya:prl1960}. Despite its small magnitude compared to the conventional exchange energy, the DMI has proven to strongly affect the spin physics of magnets. The most well-known examples are the magnetic skyrmions and helical spin phases appearing in noncentrosymmetric ferromagnets~\cite{SkyrmReview,Rybakov:prb2013,Wilson:prb2013,Meynell:prb2014,Bogdanov:zr1989}. Additionally, the relativistic exchange interactions significantly alter the spin physics at the edges of ferromagnets, where the DMI modifies the micromagnetic boundary conditions (BCs) such that magnetic surface twist states develop~\cite{Rybakov:prb2013,Rohart:prb2013,Hals:prl2017,Mulkers:prb2018,Hals:prb2019,Wilson:prb2013,Meynell:prb2014}. Notably, these boundary-driven twist states affect the uniform ferromagnetic state~\cite{Rohart:prb2013,Hals:prl2017,Hals:prb2019} as well as the shape and stability of skyrmions~\cite{Mulkers:prb2018,Rybakov:prb2013,Wilson:prb2013,Meynell:prb2014,Leonov:prl2016}. 

Here, we show that the DMI leads to even more striking phenomena at the boundaries of AFs. 
Not only will the DMI produce a boundary-driven twist state in the antiferromagnetic order parameter, but also a large surface magnetization develops due to the relativistic interactions (Fig.~\ref{Fig1}). 
Importantly, the characteristic time scale of the magnetization is in the THz regime, and it does not lead to any stray magnetic fields, except at kinks and corners of the sample where small geometry-dependent surface poles can emerge.
Thus, the surface magnetization displays the same appealing properties as the antiferromagnetic order parameter with ultrafast dynamics and negligible stray fields, while simultaneously being much easier to probe and manipulate.
Because the dynamics of the magnetization and N\' eel vector are coupled, the surface magnetization provides a new pathway for detecting and controlling the dynamics of AFs. Furthermore, we show that the magnitude and spatial variations of the observable magnetization give a direct measure of the DMI. Thus, it represents a new experimental probe for detecting antiferromagnetic DMI and, in principle, even allowing to disentangle different contributions such as homogeneous and inhomogeneous DMI.

%%%%%%%%%%%%%%%%%%%%%%%%%%%%%%%%%%%%%%%%%%%%%%%%%%%%%%%%%%%%%%%%%%%%%%%%%%%%%%%%%%%%%%%%%%%%%%%%%%%%%%%%%%%%%%%% 
% Theory: 
%%%%%%%%%%%%%%%%%%%%%%%%%%%%%%%%%%%%%%%%%%%%%%%%%%%%%%%%%%%%%%%%%%%%%%%%%%%%%%%%%%%%%%%%%%%%%%%%%%%%%%%%%%%%%%% 
We consider a collinear antiferromagnetic system with $\textit{C}_{\infty v}$ symmetry, which covers a finite region $V$.  This class of matter encompasses materials such as $\textit{K}_{2}\textit{V}_{3}\textit{O}_{8}$~\cite{Bogdanov} and $\textit{Mn}_{2}\textit{Ru}_{x}\textit{Ga}$~\cite{footnote1,AFMCnv}, as well as two-dimensional thin films lacking inversion symmetry, for example, due to the presence of an interface. The spin state of the collinear AFs is conveniently described by the N\' eel vector            $\boldsymbol{n} (\boldsymbol{r})$, which represents the direction of the collinearly ordered spins, and the vector field $\boldsymbol{m}(\boldsymbol{r})$ that characterizes the local magnetization produced by a relative canting of the magnetic sublattices~\cite{comment1}. The N\' eel vector is the order parameter field of collinear AFs. The temperature of the spin system is assumed to be far below the N\' eel temperature such that the system is close to the perfectly ordered phase. In this case, we can assume the N\'eel vector to have a fixed length of $| \boldsymbol{n} |= 1$ and only consider small magnetic fluctuations $|\boldsymbol{m}| \ll 1 $ such that $\boldsymbol{m}\cdot\boldsymbol{n}= 0$.   

The free energy of the magnetic system is to second order in the magnetization and the spatial gradients of the N\' eel vector given by the functional integral~\cite{Bogdanov}
\begin{align}
    \label{E_functional} F =&\int_{V}d^{3}r\bigg\{A \partial_{i}\boldsymbol{n} \cdot \partial_{i}\boldsymbol{n}  +\lambda\boldsymbol{m}^{2}-K_z(\boldsymbol{n}\cdot\hat{\boldsymbol{z}})^{2}+ \\
    &2\boldsymbol{d}\cdot(\boldsymbol{m}\times\boldsymbol{n}) + D\boldsymbol{[} (\hat{\boldsymbol{z}}\cdot \mathbf{n}) (\boldsymbol{\nabla} \cdot\boldsymbol{n})-(\boldsymbol{n}\cdot\boldsymbol{\nabla}) (\hat{\boldsymbol{z}}\cdot \mathbf{n}) \boldsymbol{]}\bigg\}. \nonumber
\end{align}
Here, we sum over repeated indices, $A$ ($\lambda$) parameterizes the inhomogeneous (homogeneous) exchange interaction, and $K_z$ represents the strength of the uniaxial anisotropy. We assume $K_z >0$ such that $\boldsymbol{n}$ is parallel or antiparallel with the $z$-axis in the uniform equilibrium state. The last two terms describe the homogeneous and inhomogeneous DMI. The homogeneous DMI vector is aligned along the high symmetry axis, $\boldsymbol{d}= d\hat{\boldsymbol{z}}$. Throughout this work, the DMI strengths $D$ and $d$ are assumed to be small enough to maintain a uniform antiferromagnetic phase in the interior of the sample.     

The equilibrium state of the magnetic system is found by a variational minimization of the energy functional Eq.~\eqref{E_functional} with respect to the N\' eel vector and the magnetization. Due to the normalization condition on $\boldsymbol{n}$, the variation $\delta\boldsymbol{n}(\boldsymbol{r})$ of the N\' eel vector is constrained by the condition $\delta\boldsymbol{n}(\boldsymbol{r})\cdot\boldsymbol{n}(\boldsymbol{r})=0$. Therefore, we can write the variation as  $\delta\boldsymbol{n}(\boldsymbol{r})=\boldsymbol{n}(\boldsymbol{r})\times\delta\boldsymbol{\omega}_{n}(\boldsymbol{r})$, where $\delta\boldsymbol{\omega}_{n}\in \mathbb{R}^3$ and $|\delta\boldsymbol{\omega}_{n}|\ll1$. Similarly, the local perturbation $\delta\boldsymbol{m}(\boldsymbol{r})$ of the magnetization can be written as $\delta\boldsymbol{m}(\boldsymbol{r})=\boldsymbol{n}(\boldsymbol{r})\times\delta\boldsymbol{\omega}_{m}(\boldsymbol{r})$, with $\delta\boldsymbol{\omega}_{m}\in \mathbb{R}^3$ and $|\delta\boldsymbol{\omega}_{m}|\ll1$, such that $\boldsymbol{m}$ remains orthogonal to $\boldsymbol{n}$ when the N\' eel vector is fixed. The equilibrium condition for the N\' eel vector  (magnetization) is  $\delta F/\delta\boldsymbol{\omega}_{n}=0$ ($\delta F/\delta\boldsymbol{\omega}_{m}=0$), which yields
\begin{subequations}
\begin{align}
    \label{Bulk_n}0=&\boldsymbol{n}\times\bigg\{A\boldsymbol{\nabla}^{2}\boldsymbol{n} + K_z(\boldsymbol{n}\cdot\hat{\boldsymbol{z}})\hat{\boldsymbol{z}}+ \boldsymbol{m}\times\boldsymbol{d} \\\label{Bulk_m}\notag
    &+D\boldsymbol{[} \boldsymbol{\nabla} (\hat{\boldsymbol{z}}\cdot\boldsymbol{n})\boldsymbol - (\boldsymbol{\nabla}\cdot\boldsymbol{n})\hat{\boldsymbol{z}} {]}\bigg\} , \\
    0=&\boldsymbol{n}\times\bigg\{\lambda\boldsymbol{m}+\boldsymbol{n}\times\boldsymbol{d}\bigg\} .
\end{align}
\end{subequations}
Eqs.~\eqref{Bulk_n} and \eqref{Bulk_m} determine the internal values of the N\' eel vector and the magnetization, respectively. 
Additionally, we find that the N\'eel vector satisfies the following equation at the boundaries of the sample
\begin{equation}
\boldsymbol{n}\times\bigg\{2A(\hat{\boldsymbol{\nu}}\cdot\boldsymbol{\nabla}) \boldsymbol{n} + D\boldsymbol{[}\hat{\boldsymbol{\nu}}(\hat{\boldsymbol{z}}\cdot\boldsymbol{n})-(\hat{\boldsymbol{\nu}}\cdot\boldsymbol{n})\hat{\boldsymbol{z}}\boldsymbol{]}\bigg\} = 0, \label{Boundary}
\end{equation}
where $\hat{\boldsymbol{\nu}}$ is the outer surface normal of the boundary. The boundary equation originates from the partial derivatives in the free energy functional, which give surface integrals upon variation of the N\' eel vector. Consequently, we only obtain a boundary value problem for $\boldsymbol{n}$, whereas the magnetization is controlled by the N\' eel vector via Eq.~\eqref{Bulk_m}. Note that only the inhomogeneous DMI enters the BCs, while both the inhomogeneous and homogeneous DMIs are present in the equations for the interior of the sample. 

From Eq.~\eqref{Boundary} it follows that the inhomogeneous DMI forces the antiferromagnetic order parameter $\mathbf{n}$ to develop a spin texture close to the boundaries. This effect appears even in the uniform state, in which the N\' eel vector is aligned along the $z$-axis in the bulk. In this case, the BCs imply that $(\hat{\boldsymbol{\nu}}\cdot\boldsymbol{\nabla}) \boldsymbol{n} \sim - (D/2A) \hat{\boldsymbol{\nu}}$ close to a boundary with a surface normal $\hat{\boldsymbol{\nu}}$ lying in the $xy$-plane~\cite{comment2}. Thus, the order parameter attains a modulation $\delta \boldsymbol{n} $ at the interface, where the spatial variation of the N\' eel vector is along the surface normal and on the order of $|D/2A|$. Note that this boundary-induced texture is analogous to the surface twist states reported in ferromagnets~\cite{Rohart:prb2013,Hals:prl2017}. However, the antiferromagnetic surface twist states are more complex as the spin textures additionally lead to a relative canting of the magnetic sub-lattices and a surface magnetization. This is seen from Eq.~\eqref{Bulk_m}, which implies a magnetization 
\begin{equation}\label{mag1D}
    \boldsymbol{m}=\frac{d}{\lambda}\hat{\boldsymbol{z}}\times \delta \boldsymbol{n},
\end{equation} 
for small deviations $\delta\boldsymbol{n}$ away from the uniform equilibrium state where $\boldsymbol{n} || \hat{\boldsymbol{z}}$. Note that we use the word ''surface'' instead of ''boundary'' when addressing the magnetization induced by the twist state, as this is common in the literature (e.g., chiral surface twists)~\cite{Rybakov:prb2013,Wilson:prb2013,Meynell:prb2014,Leonov:prl2016}.    

\begin{figure}
    \centering
    \includegraphics{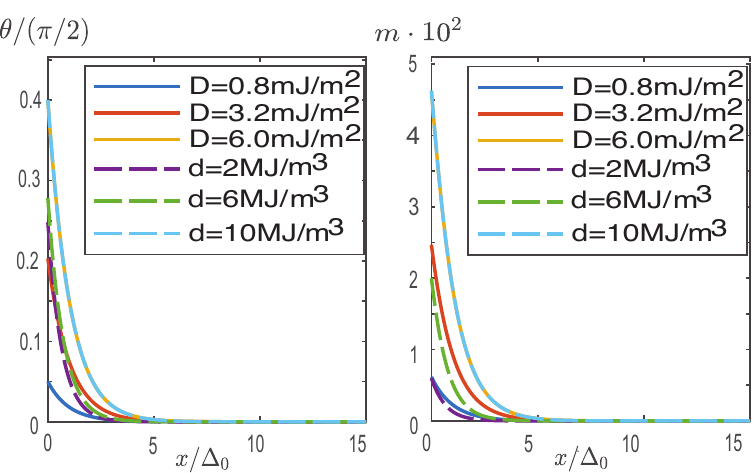}
    \caption{The tilt angle $\theta(x)$ and the magnetization $m(x)$ for different values of both the homogeneous and the inhomogeneous DMI parameters. We have used the material parameters $A= 4.8\cdot 10^{-11}$~J/m, $\lambda = 1.27\cdot10^{8}$~J/m$^3$ and $K_{z}= 1.33\cdot10^{6}$~J/m$^3$ ~\cite{Param}. When one of the DMI parameters is varied, the other is held fixed at $d=1.0\cdot10^{7}$~J/m$^3$ and $D=6.0$~mJ/m$^2$, respectively. The domain wall width $\Delta_{0}\approx9.41$ nm, is evaluated for $d=1.0\cdot10^{7}$~J/m$^3$.}
    \label{m1D}
\end{figure}

%%%%%%%%%%%%%%%%%%%%%%%%%%%%%%%%%%%%%%%%%%%%%%%%%%%%%%%%%%%%%%%%%%%%%%%%%%%%%%% 
%Results:
%%%%%%%%%%%%%%%%%%%%%%%%%%%%%%%%%%%%%%%%%%%%%%%%%%%%%%%%%%%%%%%%%%%%%%%%%%%%%%% 
To demonstrate the formation of this novel surface magnetization, we first consider a one-dimensional semi-infinite AF that extends from $x=0$ to $x=\infty$.
Due to the boundary at $x=0$, we expect the DMI to produce a twist state in the antiferromagnetic order parameter close to the edge. We parameterize the rotation by the angle $\theta$, which describes the tilting of $\boldsymbol{n}$ with respect to the $z$-axis (see inset in Fig.~\ref{Fig1}). In this case, the N\' eel vector and magnetization are given by
\begin{subequations}
\begin{align}
   \label{parn}\boldsymbol{n}(x)&=(\sin{\theta(x)},0,\cos{\theta(x)}),\\\label{par_m}
   \boldsymbol{m}(x)&=(0,m(x),0),
\end{align}
\end{subequations}
where the parametrization of the magnetization vector follows from Eq.~\eqref{mag1D}. 
Substituting Eqs.~\eqref{parn}-\eqref{par_m} into \eqref{Bulk_n}-\eqref{Bulk_m}, gives the following equations for the interior
\begin{subequations}
\begin{align}
    \partial^{2}_{x}\theta&=\sin{\theta}\cos{\theta}/ \Delta^{2},\label{1Dtheta} \\ 
    m(x)&= d\sin{\theta} /\lambda,
\end{align}
\end{subequations}
whereas Eq.~\eqref{Boundary} yields the following BC for the tilt angle
\begin{equation}
    \partial_{x}\theta (0)=-D/2A. \label{1Dthetaboundary}
\end{equation}
Here, $\Delta=1/\sqrt{K_z/A-d^{2}/\lambda A}$. Comparing the bulk equation for $\theta$ (Eq.~\eqref{1Dtheta}) to the equilibrium equation for a domain wall (DW) in an infinitely long wire~\cite{DMw}, we interpret the constant $\Delta$ as the domain wall width. From this expression we obtain the critical value for the homogeneous DMI parameter, $|d_{c}|=\sqrt{K_{z}\lambda}$, above which the antiferromagnet transitions to a weak ferromagnetic phase. For $D\geq D_{c}=(4/\pi)\sqrt{A | \lambda K_z-d^2 | /\lambda }$, modulated states are stabilized by the inhomogeneous DMI ~\cite{Bogdanov}. This defines the critical value for the inhomogneous parameter. The solution of the boundary value problem in Eqs.~\eqref{1Dtheta} and \eqref{1Dthetaboundary}  is
\begin{align}
    \theta(x)=2\arctan{(Ee^{- x/\Delta })},
\end{align}
where $E=(2A-\sqrt{4A^{2}-(D\Delta)^{2}})/D\Delta$. The solution of the tilt angle $\theta$ is displayed in Fig. \ref{m1D} for different values of the inhomogeneous DMI parameter.
 
We observe that the interior maintains a uniform antiferromagnetic state. At the edge, however, the DMI-induced BC produces a strong reorientation of the  N\' eel vector. 
For typical material parameters of AFs (given in the caption of Fig.~\ref{m1D}), the tilt angle can be as large as $\theta\approx 36^{o}$ at the boundaries (corresponding to $D=6.0$ mJ/m$^2$), thus demonstrating that the DMI significantly affects the antiferromagnetic ground state.
Additionally, in a layer of thickness $5 \Delta\sim 50$~nm close to the edge, a surface  
magnetization on the order of $M= M_s m\sim 2.3\cdot 10^4$~A/m develops  (here, we assume a saturation magnetization of $M_s= 0.5\cdot 10^6$~A/m~\cite{Param}).
This is a remarkably large magnetization that is comparable to the magnetism of ferromagnetic semiconductors such as (Ga,Mn)As~\cite{Ciccarelli:nt2015}.
Importantly, the dynamics of the surface magnetization is governed by the characteristic time scales of the N\' eel vector~\cite{comment3}.
Its typical precession frequencies are therefore in the THz regime, which differ markedly from the GHz dynamics that characterizes the magnetization dynamics of conventional ferromagnets.
This magnetization can be measured directly using imaging techniques such as magneto-optic Kerr effect microscopy and provides
a direct measure of the antiferromagnetic DMI parameters $D$ and $d$. For example, the magnitude $M$ is controlled by both the homogenous and inhomogeneous DMI, whereas the spatial variation of the magnetization is only determined by $D$. Thus, by first mapping out the value of $D$ from measurements of the spatial variations at the edges, the value of $d$ can be found from measurements of the magnitude M.  

Although the magnetization is finite in the above example, no stray field $\boldsymbol{B}$ is produced outside the sample.  
This can be seen by substituting \eqref{par_m} into the expression for the magnetic scalar potential
\begin{equation}
    \label{magpot}U(\boldsymbol{r})=\frac{M_{s}}{4\pi}\bigg(\int_{\partial V}\frac{\hat{\boldsymbol{\nu}}\cdot\boldsymbol{m}(\boldsymbol{r}^{'})}{|\boldsymbol{r}-\boldsymbol{r}^{'}|}dS^{'} -\int_{V} \frac{\boldsymbol{\nabla}^{'}\cdot\boldsymbol{m}(\boldsymbol{r}^{'})}{|\boldsymbol{r}-\boldsymbol{r}^{'}|}d\boldsymbol{r}^{'} \bigg).
\end{equation}
The demagnetizing field (i.e., the stray field) is given by $\boldsymbol{B}(\boldsymbol{r})=-\mu_{0}\boldsymbol{\nabla} U(\boldsymbol{r})$, where $\mu_0$ is the permeability of vacuum. Because the divergence of the magnetization vanishes and $\boldsymbol{m}$ is perpendicular to the surface normal $\hat{\boldsymbol{\nu}}$ at $x=0$, the magnetic scalar potential is zero for the semi-infinite AF.  
This will also be the case for circular disks, in which the N\' eel field varies along the radial direction $\hat{\boldsymbol{r}}$ ($\boldsymbol{n} \sim \boldsymbol{\hat{r}}$), such that $\boldsymbol{m} \sim \hat{\boldsymbol{e}}_{\phi}$ and $\boldsymbol{m} \cdot \hat{\boldsymbol{r}}= 0$.
 
In order to obtain a nonzero stray field, kinks or corners in the geometry of the sample are required. 
In what follows, we will illustrate this phenomenon by considering an ultra-thin film of rectangular shape with dimensions $2L_x\times 2L_y\times 2L_z$ where $L_z \ll L_x, L_y$. 
For a small DMI, the order parameter can be written as a sum of a constant part (solution for $D=d=0$) and a small perturbation that depends on the DMI: $\boldsymbol{n}(x,y)=\hat{\boldsymbol{z}}+\delta\boldsymbol{n}(x,y)$, where $|\delta \boldsymbol{n} | \ll1$. Substituting this expression into 
Eqs.~\eqref{Bulk_n} and \eqref{Boundary}, and linearizing with respect to $\delta\boldsymbol{n}$ and $\boldsymbol{m}$, we get
\begin{subequations}
\begin{align}
    \label{bulkn2Dq}&(  \partial^{2}_x + \partial^{2}_y   ) \delta\boldsymbol{n} - \delta\boldsymbol{n}/\Delta^2=0 , \\\label{bc2Dq}
    &(\hat{\boldsymbol{\nu}}\cdot\boldsymbol{\nabla})\delta\boldsymbol{n}= -(D/2A)\hat{\boldsymbol{\nu}} ,  
\end{align}
\end{subequations}
where Eq.~\eqref{bc2Dq} represents the BCs for the N\' eel vector. The expression for $\boldsymbol{m}$ follows by substitution of $\delta\boldsymbol{n}$ in Eq.~\eqref{mag1D}. 
Employing the method of separation of variables, we obtain the following solution of the above equations 
\begin{equation}
    \delta \boldsymbol{n}= - \frac{D\Delta}{2A} ( {\rm sech} (\tilde{L}_x ) {\rm sinh} (\tilde{x}), {\rm sech} (\tilde{L}_y ) {\rm sinh} (\tilde{y}), 0) .
\end{equation}
Here, we have introduced the dimensionless lengths $\tilde{L}_i = L_i/\Delta$ and $\tilde{\boldsymbol{r}}= \boldsymbol{r}/\Delta$.
As expected, we obtain a twist state in the vicinity of the sample edges, whereas the interior sustains a uniform state. 
Far away from the corner points, 
it is clear from the solution of the N\' eel vector that $\delta \boldsymbol{n} || \hat{\boldsymbol{\nu}}$ and $\boldsymbol{m} \bot \hat{\boldsymbol{\nu}}$. 
Close to the corners, however, the N\' eel vector gradually rotates, and the surface magnetization attains a component that is parallel to the surface normal $\hat{\boldsymbol{\nu}}$. The chirality of this rotation is determined by the sign of $D$. This appears in a small neighborhood of area $5\Delta\times 5\Delta$ close to each corner point. The component of the magnetization that is parallel to $\hat{\boldsymbol{\nu}}$ effectively acts as a surface pole in the magnetic scalar potential and leads to magnetic stray fields. As previously, the volume contribution to the magnetic potential vanishes, because the divergence of the magnetization is zero. 
Noting that the length scales $L_x$ and $L_y$ of the thin film typically are much larger than the domain wall width $\Delta$, the surface integral in Eq.~\eqref{magpot} can be approximated by asymptotic expansions of the appearing Laplace integrals~\cite{Bender}, which yields the following analytic expression for the stray field
\begin{align}
    \boldsymbol{B}=B_{0}\Lambda\sum_{s_{x}=\pm1}\sum_{s_{y}=\pm1}\Tilde{\boldsymbol{\nabla}}\Tilde{I}_{s_{x}s_{y}}. 
\end{align}
Here, $\Tilde{I}_{s_{x}s_{y}}= s_{x}s_{y} / \sqrt{(\Tilde{x}-s_{x}\Tilde{L}_{x})^{2}+(\Tilde{y}-s_{y}\Tilde{L}_{y})^{2}+\Tilde{z}^{2}}$ represents the surface pole at each corner,  $\Lambda=\tanh{(\tilde{L}_x)} - \tanh{(\tilde{L}_y)}$, and $B_{0}=\mu_{0}dD\Delta M_{s}\Tilde{L}_{z}/(8\pi A\lambda)$.

\begin{figure}
    \centering
    \includegraphics{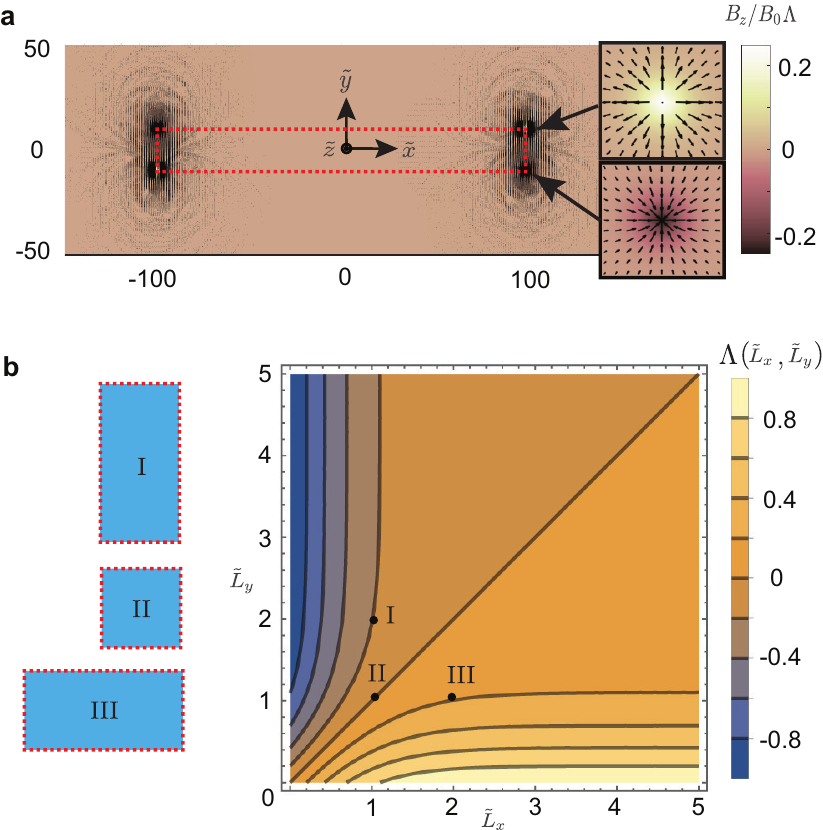}
    \caption{\textbf{a)} The dimensionless stray field ($\boldsymbol{B}/B_{0}\Lambda$) for a 2D thin film with $2\Tilde{L}_{x}= 200$, $2\Tilde{L}_{y}= 20$, and $2\Tilde{L}_{z}= 2$ at $\Tilde{z}=2.5$. The sample area is indicated by the dotted red line. The in-plane components of $\boldsymbol{B}$ are illustrated by the vectors. \textbf{b)}~The geometric factor $\Lambda$ as a function of the side lengths $\tilde{L}_x$ and $\tilde{L}_y$. The points \textrm{I}, \textrm{II}, and \textrm{III} correspond the sample geometries (i.e., the rectangles and square) illustrated on the left.}
    \label{Fig3}
\end{figure}

Fig.~\ref{Fig3}a shows the stray magnetic field for a sample of dimension $200\times20\times2$ (in units of $\Delta$) at $\tilde{z}= 2.5$ above the sample centered at $\Tilde{z}=0$ (i.e., 1.5 above the sample). This distance is within the range that, for example, a diamond nitrogen-vacancy magnetometer can provide high spatial resolution in magnetic field sensing~\cite{Bmeasure1,Bmeasure2}. Interestingly, we find that magnetic monopole fields emerge at the corners. The charges of these poles are determined by the rotation direction of the magnetization at the boundary and change signs when the sign of D is reversed. 

The magnitude of the induced stray field is mainly governed by the two parameters $B_0$ and $\Lambda$. $B_0$ is a material parameter, which is controlled by the ratio between the DMI parameters and the exchange interactions. 
The quantity $\Lambda$, on the other hand, is purely a geometric factor that is determined by the shape of the system (see Fig.~\ref{Fig3}b). The factor vanishes for quadratic samples~\cite{comment3} and reaches its maximum (minimum) value of 1 (-1) for samples in which one of the side lengths approached zero. Thus, we expect the stray field to be particularly strong (on the order of $B_0$) for nanowires and in the vicinity of structural defects such as V-notches.
For the material parameters given in the caption of Fig.~\ref{m1D} with $D=6.0\cdot10^{-3}\, \text{J/m}^{2}$, we find a magnetic field strength of $B_0= 2.3$~mT, which is experimentally detectable~\cite{Bmeasure2}.
The stray field is exponentially suppressed for samples with $\tilde{L}_x \gg 1$ and $\tilde{L}_y \gg 1$ and can be tuned negligibly small by choosing an appropriate sample geometry.  

%%%%%%Summary%%%%%%%%%%%%%%%%%%%%
In conclusion, we have demonstrated that the DMI of noncentrosymmetric AFs produces a significant surface magnetization,  
which obeys ultra-fast THz dynamics, and by engineering the device geometry one can avoid magnetic stray fields. 
We have shown that the surface magnetization provides a direct tool to measure and even disentangle different contributions of the antiferromagnetic DMI. 
Furthermore, the direct coupling of the magnetization to the change in the N\'eel vector opens a new route for manipulating AFs via the DMI-induced magnetism. 
Thus, our findings combine the key advantages of AFs (high stability, fast writing/switching speeds and ultra-high storage density) with the ones of ferromagnets (easy manipulability) and thereby contribute significantly towards AF-based information storage systems.
%%%%%%%%%%%%%%%%%%%%%%%%%%

This work received funding from the Research Council of Norway via the Young Research Talents Grant No. 286889 "Antiferromagnetic Spinmechatronics", as well as from the Deutsche Forschungsgemeinschaft (DFG, German Research Foundation) Project No. EV 196/2-1 and TRR 173 268565370 (project B12).

%%% References %%%%%%%%%%%%%%%%%%%%%%%%%%%%%%%%%%%%%%%%%%%%%%%%%%%%%%%%%%%% 


\begin{thebibliography}{99} 

%spintronics reviews:
\bibitem{Chappert:nm2007} C. Chappert, A. Fert, A. and F.N. Van Dau, Nat. Mater. {\bf 6}, 813(2007).
\bibitem{Ralph:nm2007} D.C. Ralph, M.D. Stiles, J. Magn. Magn. Mater. {\bf 320}, 1190 (2008).
\bibitem{Brataas:nm2012} Brataas, A., Kent, A. D. and Ohno, H. Nat. Mater. {\bf 11}, 372 (2012).

% AF spintronics reviews:
\bibitem{Jungwirth:np2018} T. Jungwirth, J. Sinova, A. Manchon, X. Marti, J. Wunderlich and C. Felser, Nat. Phys.  {\bf 14}, 200 (2018).
\bibitem{Duine:np2018} R. A. Duine, Kyung-Jin Lee, S. P. Parkin and M. D. Stiles, Nat. Phys. {\bf 14}, 217 (2018).
\bibitem{Gomonay:np2018} O. Gomonay, V. Baltz, A. Brataas and Y. Tserkovnyak, Nat. Phys. {\bf 14}, 213 (2018). 
\bibitem{Zelezny:np2018} J. {\v Z}elezn{\'y}, P. Wadley, K. Olejn{\'i}k, A. Hoffmann and H. Ohno, Nat. Phys. {\bf 14}, 220 (2018).
\bibitem{Nemec:np2018} P. N{\v e}mec, M. Fiebig, T. Kampfrath and A. V. Kimel, Nat. Phys. {\bf 14}, 229 (2018).
\bibitem{Libor:np2018} L. {\v S}mejkal, Y. Mokrousov, B. Yan and A. H. MacDonald, Nat. Phys. {\bf 14}, 242 (2018).

%AF dynamics using currents
\bibitem{Wadley:science2016} P. Wadley et al., Science {\bf 351}, 587 (2016).
\bibitem{Reichlova:prb2015} H. Reichlova et al.., Phys. Rev. B {\bf 92}, 165424 (2015).
\bibitem{Nunez:prb2006} A. S. N{\'u}{\~n}ez, R. A. Duine, P. Haney, and A. H. MacDonald, Physical Review B {\bf 73}, 214426 (2006).
\bibitem{Duine:prb2007} R.A. Duine, P.M. Haney, A.S. N{\'u}{\~n}ez, and A.H. MacDonald, Physical Review B {\bf 75}, 014433 (2007).
\bibitem{Gomonay:jmj2008} H.V. Gomonay and V.M. Loktev, J. Mag. Soc. Japan {\bf 32}, 535 (2008).
\bibitem{Wang:prl2008} Y. Xu, S. Wang, and K. Xia, Physical Review Letters {\bf 100}, 226602 (2008).
\bibitem{Haney:prl2008} P. M. Haney and A. H. MacDonald, Physical Review Letters {\bf 100}, 196801 (2008).
\bibitem{Gomonay:prb2010} H.V. Gomonay and V.M. Loktev, Physical Review B {\bf 81}, 144427 (2010).
\bibitem{Hals:prl2011} K.M.D. Hals, Y. Tserkovnyak, A. Brataas, Phys. Rev. Lett. {\bf 106}, 107206 (2011).
\bibitem{Manchon:prb2014} Hamed Ben Mohamed Saidaoui, A. Manchon, and X. Waintal, Physical Review B {\bf 89}, 174430 (2014).
\bibitem{Cheng:prb2014} R. Cheng, Physical Review B {\bf 89}, 081105(R) (2014).
\bibitem{Cheng:prl2014} R. Cheng, J. Xiao, Q. Niu, and A. Brataas, Physical Review Letters {\bf 113}, 057601 (2014).
\bibitem{Velkov:njp2016}H. Velkov, O. Gomonay, M. Beens, G. Schwiete, A. Brataas, J. Sinova, R. A. Duine, New J. Phys. {\bf 18}, 075016 (2016). 

%Optical AF dynamics
\bibitem{Duong:prl2004} N. P. Duong, T. Satoh, and M. Fiebig, Phys. Rev. Lett. {\bf 93}, 117402 (2004). 
\bibitem{Kimel:n2004} A. V. Kimel, A. Kirilyuk, A. Tsvetkov, R. V. Pisarev, and T. Rasing, Nature {\bf 429}, 850 (2004). 
\bibitem{Manz:np2016} S. Manz, M. Matsubara, T. Lottermoser, J. B{\"u}chi, A. Iyama, T. Kimura, D. Meier, and M. Fiebig, Nat. Photonics {\bf 10}, 653 (2016).

%DMI
\bibitem{Dzyaloshinsky:jpcs1958} I. Dzyaloshinsky, J. Phys. Chem. Solids {\bf 4}, 241 (1958). 
\bibitem{Moriya:prl1960} T. Moriya, Phys. Rev. Lett. {\bf 4}, 228 (1960).

%Skyrmions in ferromagnets
\bibitem{SkyrmReview} For reviews see A. Fert, V. Cros, and J. Sampaio, Nature Rev. Mater. {\bf 2}, 17031 (2017);  N. Nagaosa and Y. Tokura, Nature Nanotech. {\bf 8}, 899 (2013).
\bibitem{Rybakov:prb2013}F. N. Rybakov, A. B. Borisov, and A. N. Bogdanov, Phys. Rev. B {\bf 87}, 094424 (2013).
\bibitem{Wilson:prb2013}M. N. Wilson, E. A. Karhu, D. P. Lake, A. S. Quigley, S. Meynell, A. N. Bogdanov, H. Fritzsche, U. K. R{\"o}{\ss}ler, and T. L. Monchesky, Phys. Rev. B {\bf 88}, 214420 (2013).
\bibitem{Meynell:prb2014}S. A. Meynell, M. N. Wilson, H. Fritzsche, A. N. Bogdanov, and T. L. Monchesky, Phys. Rev. B {\bf 90}, 014406 (2014).
\bibitem{Bogdanov:zr1989}A. N. Bogdanov and D. A. Yablonski\u{\i}, Zh. Eksp. Teor. Fiz. {\bf 96},253-260 (1989).

%DMI & BCs
\bibitem{Rohart:prb2013}S. Rohart and A. Thiaville, Phys. Rev. B {\bf 88}, 184422 (2013).
\bibitem{Hals:prl2017}K. M. D. Hals and K. Everschor-Sitte, Phys. Rev. Lett. {\bf 119}, 127203 (2017).
\bibitem{Mulkers:prb2018}J. Mulkers, K. M. D. Hals, J. Leliaert, Milorad V. Milosevic, B. Van Waeyenberge, K. Everschor-Sitte, Phys. Rev. B {\bf 98}, 064429 (2018).
\bibitem{Hals:prb2019}K. M. D. Hals and K. Everschor-Sitte, Phys. Rev. B {\bf 99}, 104422 (2019).

\bibitem{Leonov:prl2016}A. O. Leonov et al., Phys. Rev. Lett. {\bf 117}, 087202 (2016).

%%%%%
\bibitem{Bogdanov} A. N. Bogdanov, U. K. R{\"o}{\ss}ler, M. Wolf, and K.-H. M{\"u}ller, Phys. Rev. B {\bf 66}, 214410 (2002). 

\bibitem{footnote1} $\textit{Mn}_{2}\textit{Ru}_{x}\textit{Ga}$ is a ferrimagnetic material, which in certain parameter regimes is fully compensated and effectively behaves as a noncentrosymmetric AF (see Ref.~\cite{AFMCnv}).

\bibitem{AFMCnv} N. Thiyagarajah, Y. Lau, D. Betto, K. Borisov, J. M. D. Coey, P. Stamenov, K. Rode, Appl. Phys. Lett. {\bf106}, 122402 (2015).

\bibitem{comment1} In the case of a two sublattice AF, $\boldsymbol{n}(\boldsymbol{r})=(\boldsymbol{m}_{1}(\boldsymbol{r})-\boldsymbol{m}_{2}(\boldsymbol{r}))/2$  and $\boldsymbol{m}(\boldsymbol{r})=(\boldsymbol{m}_{1}(\boldsymbol{r})+\boldsymbol{m}_{2}(\boldsymbol{r}))/2$ where $\boldsymbol{m}_{1}$ and $\boldsymbol{m}_{2}$ are the magnetic unit vectors of each sublattice.

\bibitem{comment2} Note that the twist state disappears for boundaries with $\hat{\boldsymbol{\nu}}= \hat{\boldsymbol{z}}$. 

\bibitem{DMw} A. Hubert and R. Sch{\"a}fer, \emph{Magnetic Domains} (Springer, Berlin, 1998).

\bibitem{Param} V.P. Kravchuk, O. Gomonay, D.D. Sheka, D.R. Rodrigues, K. Everschor-Sitte, J. Sinova, J. van den Brink, and Y. Gaididei
Phys. Rev. B {\bf99}, 184429 (2019).

\bibitem{Ciccarelli:nt2015}C. Ciccarelli, K. M. D. Hals, A. Irvine, V. Novak, Y. Tserkovnyak, H. Kurebayashi, A. Brataas, and A. Ferguson,  Nature Nanotech. {\bf 10}, 50 (2015). 

\bibitem{comment3} See Supplemental Material.

\bibitem{Bender} C. M. Bender and S. A. Orszag, \emph{Advanced Mathematical Methods for Scientists and Engineers} (Springer-Verlag, New York, 1999).

\bibitem{Bmeasure1} P. Maletinsky, S. Hong, M. Grinolds et al. Nature Nanotech {\bf 7}, 320 (2012).

\bibitem{Bmeasure2} J.L. Webb, J.D. Clement, L. Troise,  S. Ahmadi, G.J. Johansen,  A. Huck, and U.L. Andersen 
Appl. Phys. Lett. {\bf 114}, 231103 (2019).

\end{thebibliography}
\end{document}